\documentstyle[astron,12pt]{l-aa}
\newcommand{\EQ}{\begin{equation}}
\newcommand{\EN}{\end{equation}}
\newcommand{\EQA}{\begin{eqnarray}}
\newcommand{\ENA}{\end{eqnarray}}

\input psfig
\begin{document}
\thesaurus{12.03.2, 12.12.1}
\title{ Pure Luminosity Evolution Hypothesis for QSOs: $\qquad$$\qquad$
From Luminosity Functions to Synthetic Catalogues}
\author{G. Mathez$^1$, L. Van Waerbeke$^1$, Y. Mellier$^1$, H. Bonnet$^1$, M. Lachi\`eze-Rey$^2$}
\offprints{G. Mathez}
\institute{  $^1$ Laboratoire d'Astrophysique de Toulouse, URA285.
Observatoire Midi-Pyr\'en\'ees
14, avenue Edouard Belin F-31400 - Toulouse
$\;\; ^2$ CEN Saclay Service de Physique Th\'eorique F-91191 Gif Cedex
}

\date{ Received ; accepted }
\maketitle
\markboth{G. Mathez {\it et al.}: Synthetic QSO catalogues}{}
\begin{abstract}
This paper describes the construction of
realistic extragalactic Monte--Carlo catalogues, 
aimed at comparing the behaviour of cosmological tests versus 
input parameters. QSO catalogues are built on a luminosity 
function derived from data through a certain choice of any cosmological
and evolutionary model, and through suitable computation of 
individual maximum volumes in complete (but magnitude--
and redshift--limited) samples. Details of this computation are examined 
for the case of strong luminosity evolution. 
The values of the evolution parameter 
are derived for various cosmologies,
corresponding to $<V/V_{max}>=1/2$ 
in the sample of 400 Ultra--Violet
Excess (UVX) QSOs \cite{boyle90}.
The various luminosity functions 
are compared, both for the whole sample  
and in redshift bins. A characteristic evolution
time is defined and computed,
depending strongly on the cosmology, but practically
constant when expressed in terms of the age of the universe. 
Algorithms are given for producing catalogues
based on the null hypothesis that the objects 
are uniformly distributed in volume but suffer pure luminosity evolution.
The only input based on real data is the luminosity function,
but this requires neither a redshift nor 
an apparent magnitude histogram. Simulated redshift and absolute magnitude
histogrames are compared to real data.
\end{abstract}
\section {Introduction}

Various judicious tests have been developed 
to discriminate between different cosmological models, 
all requiring large and complete samples of
extragalactic objects (quasars or galaxies). 
Unfortunately, the available samples are 
magnitude limited, selection effects induce poorly known systematic 
biases in magnitude and redshift, and extragalactic populations 
undergo strong evolution over a Hubble time.
This paper describes the construction of extragalactic synthetic
catalogues with various parent cosmologies, evolution laws and 
simulated observational biases, with the goal of
applying them to the cosmological test described in Paper I 
\cite{van95}. A preliminary study of the luminosity 
function of the real population and of
its evolution in various cosmologies 
is necessary for such a construction; so it is
also a by product of the present paper.

The two main extragalactic populations, quasars and galaxies, 
may be simulated: we focus on quasars in the following since
they are more distant and {\it a priori} more
suited to cosmological tests. A quasar being by definition 
stellar in appearance, the pair of apparent magnitude--redshift 
$(m,z)$ is sufficient to characterize it with respect to current
cosmological tests.
The result of the simulation is thus a list 
of apparent magnitudes and redshifts, {\it not} an image.

The observed population suffers strong evolution, 
as evinced by either the standard
$<V/V_{max}>$ analyse or direct derivation of
the luminosity function in redshift bins. 
Pure Luminosity Evolution (PLE) seems to
be the best model \cite{boyle88}, but Pure Density Evolution 
(PDE) or mixed models are not
rejected; and the question remains open for redshifts greater than
$2.2$ \cite{shaver95}.
Paper I describes a cosmological test which accounts for such
evolution.

Quasars also suffer various biases in their redshift distribution
and a strong variability in luminosity. 
Our initial aim was to construct synthetic quasar catalogues capable of
quantifying the behaviour of the test as a function of the various parameters:
\begin{itemize}
\item sensitivity to the cosmological parameters;
\item  photometric uncertainties and strong quasar variability;
\item  redshift uncertainties and biasing in certain redshift ranges;
\item  catalogue size and redshift
limitations, leading to a preferred observational strategy for the
construction of quasar samples;
\item  this list is certainly not exhaustive.
\end{itemize}
Last, but not least, most cosmological tests using QSOs rely on some
basic assumption concerning their evolution. As a consequence, their result in
principle applies only within this framework. With Monte--Carlo catalogues,
it is possible to test the uncertainty 
induced by any false hypothesis on evolution
by making cross--tests between different assumed evolutions: 
this may concern the nature of the evolution, either
Density or Luminosity Evolution (DE/LE), as well as its functional
form, PoWer law or EXPonential (PWLE/LEXP). 
The present paper addresses only the case of PLE. 
PDE will be the subject of a forthcoming paper.

The paper is organized as follows: in Section 2
the basic method is exposed.
After a brief review of the cosmological frame, the 
computation of volumes and maximum volumes
is detailed in Section 3.
In Section 4 the sample of Boyle is 
used to derive the necessary LF.
In Section 5 the evolutionary parameters and characteristic times 
found in various cosmological models are compared. In Section 6
a general algorithm to construct artificial QSO samples is
derived from our working hypotheses. 
The redshift and magnitude histogrames of synthetic catalogues are
examined.

\section{Basic Method}

Previous numerical simulations have been performed 
in a similar way, essentially
aimed at producing catalogues of galaxies.
Chokshi {\it et al.} \cite*{chokshi88} simulate galaxies in 
clusters uniformly distributed in space.
In each cluster, about 1,000 galaxies are drawn, 
all at the same redshift. The magnitudes are sorted according 
to the Schechter Luminosity Function
(LF), brighter than a certain cut-off which of course depends on 
the cluster redshift.
Yee \cite*{yee91} first sorts the apparent magnitude according to 
the observed distribution, the
redshift is then computed in 
bins of apparent magnitude according to a certain Luminosity Function 
(LF) based on a given, fixed, cosmological and evolutionary model.
Our method is slightly different: first, 
a functional form for the evolution is chosen, 
and the cosmological 
parameters are varied on a certain grid. 
There is then a unique evolution parameter ensuring
the condition $<V/V_{a}> = 1/2$ for a given, actual,
sample, $V_{a}$ being the available
comoving volume according to Avni {\it et
al} \cite*{avni81}. The zero--redshift LF $\Phi $ is fixed by such a procedure, 
and fitted according to some previous analytic choice 
for the LF. Once the evolution of the LF 
is known, $V_{a}(M_0)$ may be computed in terms of the zero--redshift 
absolute magnitude $(M_0)$. The resulting probability distribution function 
{\it in the sample} of
absolute magnitude is then $\Phi (M_0)\times V_{a}(M_0)$. 
It is easy to sort the absolute magnitude of object $i$ according to this PDF.
A second parameter, either the apparent magnitude or the redshift,
 is needed to define a QSO. Given the absolute
magnitude $M_0$ and the cosmological and evolutionary model, 
$V_a(M_0)$ may be computed. A second random drawing according to a uniform
distribution gives a number $x \in [0,1]$, and the redshift of 
object $i$ may be computed from the condition: $V=x \times V_{a}(M_0) $. 
$(m,z)$, the apparent magnitude--redshift 
pair, is then derived from these two variables. For each cosmological model, 
the procedure is repeated until the desired catalogue size is reached.

The null hypotheses are: that PLE applies, that the 
luminosities are distributed according
to the observed LF and that, according to 
the Cosmological Principle, the QSOs are 
uniformly distributed in volume on Gpc scales, so that
the variable $V/V_{a}$ is uniformly spread over $[0,1]$. From the LF
we are able to draw
the two variables $M_0$, the zero--redshift absolute 
magnitude, and $x$; and from the assumption of uniformity we draw 
the $V/V_{max}$ ratio. 
Within this framework, particular 
attention must be paid to the computation of individual $V/V_{a}$.

\section{Theoretical background}
\subsection{Cosmology }

The universe is described by a Friedman-Robertson-Walker model which is
defined by three independent fundamental parameters (calculated at
$z=0$): the matter
density $\Omega_{m}$, which may include a non-baryonic component, the
cosmological
constant $\Omega_{\Lambda}$, and the Hubble parameter $H_0$. The value
of $H_0$ is very important for the creation of synthetic catalogues because it
scales the distance and the absolute luminosity of the objects.
Nevertheless, $H_0$ has no influence on individual $V/V_{a}'s$, which
are scale free. The other known cosmological parameters are
related to these three parameters.

No arbitrary choice of a preferred cosmological model is made.
General formulae
for the distance and the volume are given in Caroll {\it et al.}
\cite*{caroll92}. 
We define the conformal coordinate $\chi (z)$ as:

\EQA
\chi (z)&=&\, 2\, |\Omega_K|^{1/2}\,\,   \times \nonumber\cr
&&\nonumber\cr
&& \quad \,  \int_0^z {dz'\over ((1+z')^2(1+\Omega_m z')-z'(2+z')\Omega_{\Lambda})^{1/2}} \quad ;
\label{integ}
\ENA
\noindent
The volume out to redshift $z$ is written:

\EQ
V(z)=\left({c\over {H_0}}\right)^3{1\over 2\Omega_K|\Omega_K|^{1/2}}\left[ 
sinn(\chi(z))-\chi(z) \right],
\label{volu}
\EN
\noindent
where $\Omega_K=1-\Omega_m-\Omega_{\Lambda}$ is the curvature, $sinn$ 
is $sin$ if $\Omega_K<0$ and $sinh$ if $\Omega_K>0$.
This equation is very useful for numerical computation because
it does not require inverse trigonometric functions. Contrary to 
what happens in a closed universe, with the expressions given, e.g.,
in Caroll {\it et al.} \cite*{caroll92},
Eq.(\ref{volu}) is single valued and there is no discontinuity at 
the value $\chi =\pi $, as shown in Mathez {\it et al.} \cite*{mathez91}.

\subsection{Evolution}
The quasar population resides in an extremely young universe, 
and it has evolved up until the present epoch. 
It is likely that individual QSOs do not live for a
long period of time,
so we speak rather about the evolution in a statistical sense.
In any case, assuming a PLE model for quasars
amounts to assuming that the fraction of active quasars is
constant with epoch, their luminosity decreasing with increasing
time, and that they are in sufficient number so that 
we may apply this average evolutionary law to individual quasars in
order to derive maximum redshifts. The absolute
magnitude at redshift $z$ is given by:
\EQ
M(z)=M_0-2.5log_{10}(e(z)),
\label{evolu}
\EN
\noindent
where $M_0$ is the absolute magnitude at $z=0$, and $e(z)$ is the
evolutionary law: 
\EQ
e(z)\equiv (1+z)^{k_L}
\label{pwle}
\EN
for a Power Law Luminosity Evolution (PWLE), and
\EQ
e(z) \equiv exp\left(k_L t(z)/H_0\right) = exp\left(t(z)/\tau \right)
\label{lexp}
\EN
for an Exponential Luminosity Evolution (LEXP); $k_L$ is the evolutionary
parameter, $t(z)$ the look-back time and $\tau $ the 
characteristic evolutionary time. Strictly speaking, 
as quoted by Bigot and Triay \cite*{bigot91}, this
evolution law has only a statistical sense and 
does not apply to individual QSOs. However, to compute each $V/V_{a}$ value, we 
fictitiously move each QSO toward high redshifts, and it is necessary to apply 
this evolution law to each QSO.

Assuming a power law spectrum $f_\nu=\nu ^{-\alpha}$ with a spectral
index $\alpha=0.5$, the Mattig relation relates the observables
$(m,z)$ of a quasar to $M_0$, its absolute magnitude at $z=0$:

\EQ
m_{M_0}(z)=M_0-5+2.5log_{10}\left({d_L^2(z) K(z)\over e(z)}\right);
\label{mattig}
\EN
\noindent
here $K(z)= (1+z)^{(\alpha-1)}$ is the K-correction.

\begin{figure*}
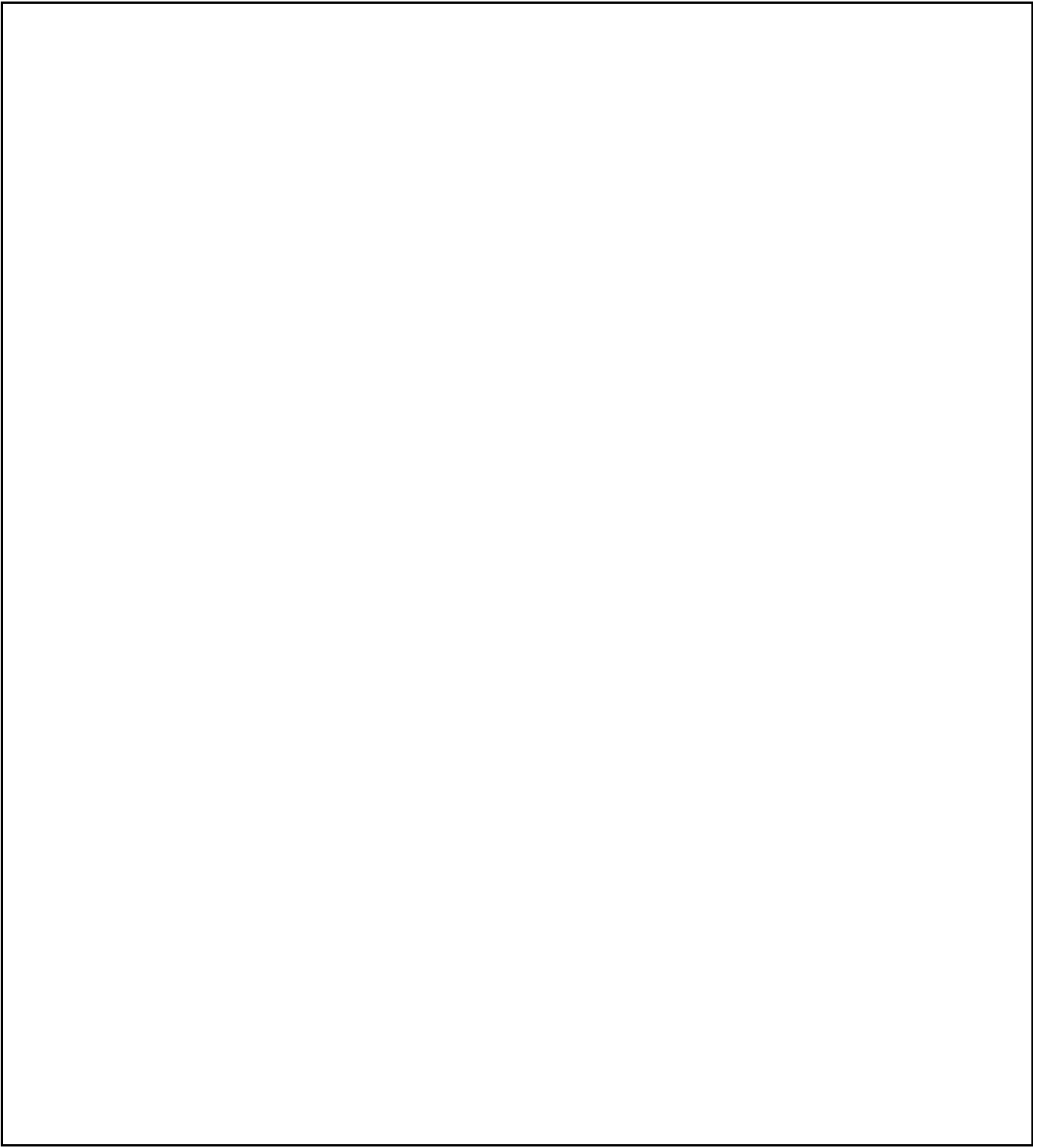

\picplace{20cm}
\caption{%
Four possible $apparent\; magnitude-redshift$
plots: monotonicaly increasing, one maximum,
both one maximum and one minimum, and monotonicaly decreasing.
The evolutionary tracks
are drawn according to the cosmology and exponential luminosity
evolution as described by the top labels. The $m=m(z)$ curves (solid lines)
correspond to bins of constant width in absolute magnitude.
The horizontal dashed
line represents a constant limiting magnitude $B=21$; the dashed $m(z)$
line corresponds to the faint separator ${\cal S}_2$ between
magnitude-limited and redshift-limited quasars (see text).
Depending on the parameters (cosmology and evolution),
the same quasar may be
either magnitude-- or redshift--limited.
In models where $m(z)$ shows a maximum (for example in Fig. 1b),
for quasars lying between $\cal S$ and $B=21$ there exists a
redshift slice where $m(z)$ is above $B=21$. The corresponding volume
must be subtracted from the available volume $V_2-V_1$.
Quasars of the complete
Boyle {\it et al.} (1990) sample are superimposed.
}
\end{figure*}

The evolutionary tracks $m(z)$ on the Hubble diagram
are shown in Fig. 1 for a given set of quasars in
four different cosmologies and evolutionary histories.
The curves appear to be quite
different. In particular the quasars in Fig1.b,c,d become brighter and 
brighter as the redshift increases, because the evolution 
dominates the cosmological dimming at large $z$.

\subsection{Distribution of the QSOs in the Hubble plane}

Available complete QSO samples usually have a well defined
limiting magnitude, towards faint magnitudes. Due to saturation
effects (in the photometry or in the spectroscopy), a bright limiting
magnitude may also apply.

All UVX QSOs are limited to redshifts $z\le 2.2$, precisely because of
their selection as objects with a UV excess, a standard way to detect 
quasars in optical astronomy.
As the redshift decreases (below $z=0.3$), there 
is another limitation because quasars become spatially 
resolved and can no longer be classified as stellar objects.
These redshift limitations prevent the use of maximum (respectively minimum) 
redshifts outside the limiting redshifts of the sample.
Assume that we have a QSO sample in the magnitude 
and redshift ranges $[m_{1},m_{2}]$ and $[z_1,z_2]$, respectively. 
For $UVX$ samples we would have $z_1=$0.3 and $z_2=$2.2.

We define $\Phi (M,z)$ as the LF at redshift $z$ and at
absolute magnitude $M$. However, for the case of PLE it
can be written $\Phi (M(z))$; then all luminosity
functions at different redshifts differ by an additive constant along 
the absolute magnitude axis. This
allows us to work with the LF defined at a fixed epoch
($z=0$) where $\Phi (M(z=0))=\phi (M_0)$, whatever the evolutionary law
\cite{kassiola91}. The number
of objects in the range of absolute luminosity $[M_0,M_0+dM_0]$ and
redshift $[z,z+dz]$ writes:

\EQA
d^2N=\phi (M_0) \, dM_0  \, H(m_2-m)  \, H(m-m_1)\;\; \times \hfill \nonumber\cr
\hfill  \, H(z_2-z)  \, H(z-z_1) \, {dV\over dz},  \, \, \, dz \;\;\; ,
\label{dene}
\ENA
\noindent
with the following notations:
\begin{itemize}
\item $dV$ is the comoving volume element of 
the shell of redshift $z$ to $z+dz$ in the
solid angle $\omega $ of the survey;
\item $\phi (M_0)$ is defined above;
\item The four Heaviside functions are the selection function,
ensuring that apparent magnitudes and redshifts remain inside 'box
$\cal B$', or $[m_1,m_2]\otimes[z_1,z_2]$, 
the limitations due to the selection criteria.
\end{itemize}

We now define ${\cal D}$ as the domain in the Hubble
plane where all quasars are redshift--limited.

Let $z_{top}$ (respectively $z_{bot}$) be the redshift in 
the range $[z_1,z_2]$ 
at which the evolutionary track passes through 
the maximum (minimum) of the $m(z)$ curve. $z_{top}$ ($z_{bot}$) is
the same for all quasars, since
their $m(z)$ curves are parallel, differing by 
an additive constant equal to the absolute magnitude. Consider a given
absolute magnitude $M_0$, and
let $m_{M_0,top}$ ($m_{M_0,bot}$) be the corresponding 
faintest (brightest) apparent 
magnitude in box $\cal B$:
\EQA
m_{M_0,top}&=\sup_{z\in [z_1,z_2]}\left[ m_{M_0}(z) \right]&= m_{M_0}(z_{top});\nonumber\cr
\\
m_{M_0,bot}&=\inf_{z\in [z_1,z_2]}\left[ m_{M_0}(z) \right]&= m_{M_0}(z_{bot}).
\label{topbot}
\ENA
$m_{M_0}(z)$ is the apparent magnitude corresponding to the
absolute magnitude $M_0$ at redshift $z$.
$z_{top}$ is equal to $z_2$ in the case of little or no luminosity
evolution. Consider the $m(z)$ curve 
corresponding to $m_{top}=m_{2}$: it is unique; let us 
call it the 'faint separator'
${\cal S}_2$. There is also a 'bright separator' -- the unique curve ${\cal S}_1$
crossing $(z_{bot},m_{bot}=m_1)$. ${\cal S}_2$ is drawn on Fig. 1.
Domain ${\cal D}$ is simply 
the domain located {\it below } ${\cal S}_2$ and {\it above} ${\cal
S}_1$. 
Let $M_0^{S_i}$ be the two absolute magnitudes corresponding to
the curves ${\cal S}_i, \; (i=1,2)$. The domain ${\cal D}$ simply 
consists of that part of the Hubble plane brighter than $M_0^{S_2}$ and
fainter than $M_0^{S_1}$. We call $\cal D^*$ the complement of $\cal D$
with respect to the box $\cal B$. We call $\cal D$ and $\cal D^*$
quasars the quasars located in each of these domains.

%%%

\subsection{The $V/V_{max}$ test}

Since our method relies on $V/V_{a}$ values, it is necessary
to review the $V/V_{a}$ computation, particulary in the case of strong
evolution.
Consider a quasar in the sample with $(m,z)$ as its representative point 
in the Hubble diagram, and define the following quantities:
\begin{itemize}
\item $M_0$: the quasar absolute magnitude at epoch $z=0$;
\item $z_{max}(M_0,m_i) \;(i=1,2)$: the redshift(s)
where the quasar has an apparent magnitude $m_i$;
\item $V_{max}(M_0,m_i)$: the corresponding comoving volume(s);
\item $V_{a}$: the available volume;
\item $V_i \; (i=1,2)$: the comoving volumes enclosed by 
the two limiting redshifts $z_i$.
\end{itemize}
The various possible shapes of the evolutionary track (Fig. 1) are 
crucial for understanding the $V/V_{a}$ test. 
The quasar may leave the sample in one of two ways:
either its magnitude leaves the range $[m_1,m_2]$, or its
redshift leaves the range $[z_1,z_2]$. The available volume
$V_{a}$ does {\it not} reduce to $V_{max}(M_0,m_2)$:
it is the total volume $V_2-V_1$, minus $\sum V_{out}$, 
the sum of the volumes of all slices of the universe where
the evolutionary track is outside box $\cal B$,
so that it prevents the quasar from remaining in the sample.

Assuming that volumes are randomly distributed over the range $[V_1,V_2]$,
the variable uniformly distributed over $[0,1]$ is the ratio $x$ defined
below.  The number of quasars in the redshift range $[z_1,z_2]$ is:

\EQA
N&=&\int_{(M_0)_1}^{(M_0)_2}dM_0 \Phi (M_0) \int_{sup(z_1,z_{max}(M_0,m_1)}^{inf(z_2,z_{max}(M_0,m_2))}dz{dV\over dz} \nonumber\cr
\\
&=&\int_{(M_0)_1}^{(M_0)_2}dM_0  \Phi (M_0)V_{a}(M_0),
\label{numb}
\ENA
where:
\EQA
V_{a}(M_0)&=&inf(V_2,V_{max}(M_0,m_2)) \;\; - \cr
&& sup(V_1,V_{max}(M_0,m_1))\;-\;\sum V_{out} .
\label{available}
%\\
\ENA
\noindent
Note that for $\cal D$ quasars
$ \; inf(z_2,z_{max})=z_2$ and $sup(z_1,z_{min})=z_1$ {\it
simultaneously}. 

It is straightforward to find the new variable $x$: 

\EQ
x=(V-V_{sup}-\sum V_{out})/(V_{inf}-V_{sup}-\sum V_{out}).
\label{nvar}
\EN
\noindent
Indeed, one shows that the mean of the ratio $x$ over the sample is
$1/2$:

\EQA
<x>&=&{1\over N} \int_{(M_0)_1}^{(M_0)_2}dM_0  \Phi (M_0)\;\;\times \cr
&& \int_{V_{sup}}^{V_{inf}} dV
{(V-V_{sup}-\sum V_{out})\over (V_{inf}-V_{sup}-\sum V_{out})} \nonumber\cr
&&\nonumber\cr
&&\nonumber\cr
&=&1/2, \qquad \forall \Phi , \forall (M_0)_1,  \forall (M_0)_2;
\label{undemi}
\ENA
\noindent
moreover, since $dV=V_{a}(M_0) dx$, we have:

\EQA
dN&=&\int_{(M_0)_1}^{(M_0)_2} dM_0 \Phi (M_0)V_{a}(M_0) \; dx\nonumber\cr
&&\nonumber\cr
&&\nonumber\cr
&=&N \;\; dx,
\label{unif}
\ENA
\noindent
which characterizes a PDF of $x$ which is uniform 
over $[0,1]$. For a redshift--limited sample over $[z_1,z_2]$,
taking $V_{sup}=0$ overestimates $V/V_{a}$ and taking $V_{inf}=V_{a}$
underestimates $V/V_{a}$. In either case, $<x>$ cannot be equal to
$1/2$ {\it even in the correct cosmology and with the correct evolutionary 
law.}

Varying the shape of the evolutionary
track changes the expression of the variable $x$.
There may exist zero, one or several maxima of $m(z)$, or
the quasar may become fainter than $m_2$ at low redshift; in fact
there may be several roots of the equation $m(z)=m_2$.
Fig. 1 shows various curves $m(z)$ for various cosmologies and 
evolutionary histories, in order to illustrate the existence of
redshift ranges, lying inside the range $[z_1,z_2]$, 
but for which $m(z)$ is {\it not} in the range $[m_1,m_2]$. Such ranges
must be excluded from the volume computation.

\section{ Luminosity Functions in real samples}

The LF is not a direct observable -- on the contrary it depends on the
choice of the cosmological model and of the evolutionary law.
For our purpose, it is
necessary to be realistic in reproducing the data; in particular, the artificial
catalogues must have the same LF as the real sample 
whatever the choice of their parent cosmological model. 

As discussed in Kassiola and Mathez, \cite*{kassiola91}, 
given the cosmology and the
evolution, there are two ways to derive the LF from data:
firstly one can construct the {\it Global Luminosity 
Function} (GLF), by shifting the absolute magnitudes of the whole sample 
to the present epoch $z=0$, according to the evolution law, and secondly
one can compute the {\it Restricted Luminosity Functions} (RLFs), i.e.
the LFs in different redshift bins.
A necessary condition for PLE to apply is that the LF in all redshift 
bins differ by a simple shift along the absolute magnitude axis. 
To do this, one has to
compare the RLFs. Assuming that PLE is correct, the reason for choosing 
the GLF to construct the catalogues is to avoid the noise created by
binning the data in redshift. In Section 4.3, we test the 
PLE hypothesis by comparing of the RLF and the GLF.

\subsection{ Real data: the Boyle {\it et al.} sample }

The complete sample of $UVX$ quasars of Boyle {\it et al.} (1990) is used. 
It contains
383 quasars, if we exclude the Narrow Emission Lines (NL) and 
the redshifts outside the range $[0.3,2.2]$.

\subsection{The Global Luminosity Function and its derivation}

We now consider the computation of the GLF for any redshift--and magnitude 
limited--QSO sample.
The sample is binned in absolute magnitude $M_0$ with bin width $\Delta M_0$.
Let $N_i$ be the QSO number in the $i^{th}$ absolute magnitude bin.
The GLF in bin $i$ is given by:
\EQ
\phi _{obs}^i (M_0)=\sum_{j=1\atop M_{0j}\in B_i}^{N_i} \left(1/ V_a^j\right),
\label{fctlum}
\EN
\EQA
B_{i}= \left[M_0-{\Delta M_0\over 2},M_0+{\Delta M_0\over 2}\right],
\ENA
\noindent
with an error on $\phi$:
\EQ
\sigma_\phi^i=\left(\sum_j^{N_i} {V_a^j}^{-2} \right)^{1/2}.
\label{erreur}
\EN
\noindent
For the case of the
Boyle sample, none of the eight fields intersect; thus, the
available volume follows directly from Eq. (8):
\EQA
V_a&=&\sum_k^{N_{fields}} \omega_k \;\; \times \cr
&& \left[ inf(V_{max}^k,V_2)-sup(V_{min}^k,V_1)- \sum V_{out} \right],
\label{vola}
\ENA
\noindent
where $\omega_k$ is the solid angle of the $k^{th}$ field.

The GLF looks similar for all of the cosmologies we tried. 
Special attention has been paid to some of the cosmological models 
(e.g. $\Omega =0, \Lambda=0;\; \Omega =1, \Lambda=0;\; \Omega =0, 
\Lambda=0.8$), by various authors which 
found that the LF is well fitted by either a single 
\cite{marshall83} or a double \cite{boyle88} power--law. We adopted
the double power law--model in all cosmologies and for the two functional
forms of evolution:

\EQ
\phi _{model} (M_0)={\phi^\star\over
10^{0.4(M_0-M_\star)(\alpha+1)}+10^{0.4(M_0-M_\star)(\beta+1)}};
\label{funcform}
\EN
\noindent
$M_\star$ is the characteristic absolute magnitude of the two power--law
distribution, i.e. the knee of the distribution. Absolute magnitudes
for the quasars are computed according to Eq. (\ref{mattig}). The
evolution parameter $k_L$ is computed by setting $<x>=1/2$ for the
whole sample. Such a procedure differs from current work 
\cite{marshall83,boyle88}, in which $k_L$ is
fitted like the other parameters. The free parameters $\phi^\star$, $M_\star$, $\alpha$, and
$\beta$ are derived by a $\chi^2$ minimization of the binned GLF:

\EQ
\chi^2=\sum_{i=1}^{N_{bin}} \left({\phi_{obs}^i-\phi_{model}\over
\sigma_{\phi_{obs}}^i} \right)^2,
\label{chi2}
\EN
\noindent
where $N_{bin}$ is the number of absolute magnitude bins.
An important difference with other LF computations 
is that we have only four free parameters. 

%\begin{figure*}
%\vskip -1truecm
%\centerline{ \hfill %%
%\psfig{figure=lumpwl.ps,width=18. cm}
% \hfill %%
%}
%\caption[]{Global Luminosity Functions in four cosmologies, for a power
%law luminosity evolution. From the thicker to the larger line are models
%$d,c,b,a$.}
%
%\vskip -1truecm
%\centerline{\hfill %%
%\psfig{figure=lumexp.ps,width=18. cm}
% \hfill %%
%}
%\caption{Same as above, exponential luminosity evolution. From the
%thicker to the larger line are models $d,c,b,a$.}
%\vfill
%\end{figure*}
\begin{figure}
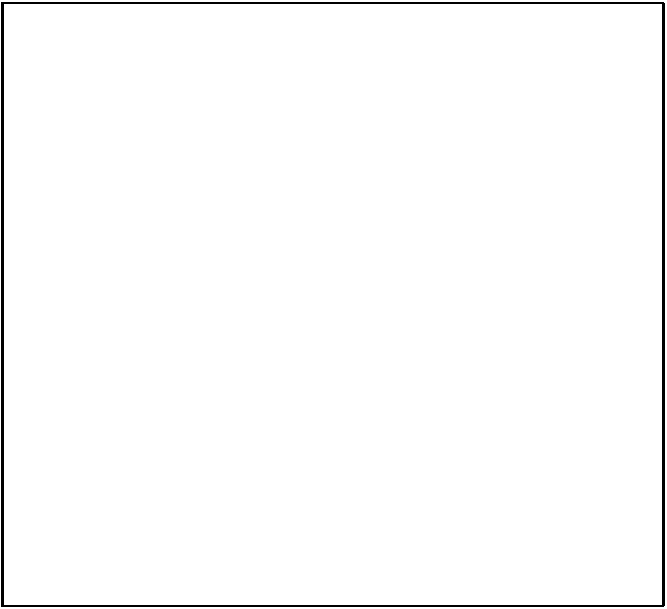

\picplace{8cm}
\caption{Global Luminosity Functions in four cosmologies, for a power--law
luminosity evolution. The models $a,b,c,d$ of Table 1 progress from the
larger to the thicker line.}
\end{figure}
\begin{figure}
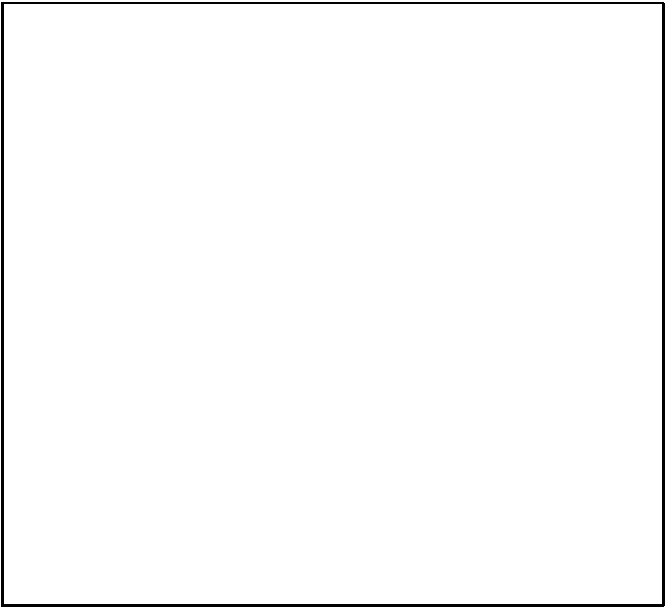

\picplace{8cm}
\caption{Same as above, exponential luminosity evolution. The models 
$e,f,g,h$ of Table 1 progress from the larger to the thicker line.}
\end{figure}

Table 1 and Figs. 2 and 3 show a panel of GLFs computed for four different
cosmologies, scaled to $H_0=50 \; km s^{-1} Mpc^{-1}$. 
The general behaviour of the GLF is identical for the
exponential and power--law evolutions.
The GLF is enlarged and QSOs are brighter for a $\Lambda $--dominated
universe. This is due to the fact that, at fixed redshift, the distance in
this type of model is greater than in other models,
and therefore the absolute luminosities are larger. The effect is in
the opposite sense
for matter--dominated universes where the objects become fainter and the GLF
narrower. The knee of the luminosity distribution evolves according to
the preceding remarks. The faintest, shortest and densest GLF is obtained
for $\Omega=1$ and $\Lambda=0$.
The evolution becomes weaker in $\Lambda$-dominated universes but has
a more standard value ($k_L$ around 3--4 in PWLE models) 
if $\Omega$ is greater than 0.4, regardless of the value of $\Lambda$.

Despite our different LF computation method, we find results quite similar to
those of Malhotra {\it et al.} \cite*{malhotra95} and Boyle {\it et al.} 
\cite*{boyle90}.

\subsection{ The Restricted Luminosity Functions }

Restricted LFs (RLFs) have been computed from the Boyle sample
in three different redshift bins
($[0.3,1.0]$, $[1.0,1.7]$, $[1.7,2.2]$).
We are interested in checking whether or not the
PLE hypothesis is confirmed. If this is the
case, any RLF may be deduced from the GLF by
a simple shift along the absolute magnitude axis according to
the evolution law (Eq. \ref{evolu}). Since we have computed 
the GLF in the last
Section, we are able to compare our RLF to the GLF.
This is done by a $\chi^2$--procedure between the GLF and the three RLFs:
\EQ
\chi^2=\sum_{i=1}^{N_{bin}} \left({\phi_{GLF}-\phi_{RLF}^i\over
\sigma_{\phi_{RLF}}^i} \right)^2.
\label{chi2}
\EN
Here, $\phi_{RLF}^i$ and $\sigma_{\phi_{RLF}}^i$ refer to the RLF, computed
from the data in the limited redshift bins, and $\phi_{GLF}$ is shifted to
this redshift range.

The result is shown in Table 1
for four cosmologies and the two evolutionary models. The probabilities
$P(>\chi^2)$ are high, and, except for one case 
(LEXP, $\Omega_{mat}=\Lambda=1$),
the test succeeds to better than the 15\% confidence level. 
This result is in condradiction to those of a similar
test done by Kassiola and Mathez (1991), which found strong discrepancies
between the GLF and the RLF. However, they interpret these discrepancies 
as the
lack of homogeneity of their composite catalogues.
Since our results are consistent with PLE (favoured by Boyle {\it et al.},
1987), we use this hypothesis to construct our synthetic catalogues.

\section{ Evolution characteristic time in various cosmologies }
In the exponential model, the characteristic evolution time $\tau $
enters explicitely in Eq. (\ref{lexp}), the expression for $e(z)$.
This is not the case for the power law model -- but a characteristic time
may be defined as the look--back time to which all luminosities were
higher by a factor $e$:
%\EQA
%\tau &=& t(z_k).\nonumber\cr
%\\
%z_k &=& exp\left(1/k_L\right) - 1
%\label{tau}
%\ENA
\EQ
\tau = t\left(z_k = exp(1/k_L) - 1\right) .
\label{tau}
\EN
Table 2 (3) gives the evolutionary parameters $k_L$ for a grid of
cosmological parameters in PWLE (LEXP).
Table 4 (5) gives the corresponding PWLE (LEXP) characteristic times, 
computed according to 
Eqs. (\ref{tau}) and (\ref{lexp}), respectively.
These times are in units of the Hubble time $H_0^{-1}$.
As is well known, the quasar evolution time is of the order of 
$H_0^{-1}/(7.7\pm3.5)$ for exponential evolution
\cite{mathez76,marshall83,boyle88},
strongly dependent on the cosmological model. For power--law
evolution, the characteristic time is 
$H_0^{-1}/(3.7\pm 0.6)$.

Tables 6 and 7 give the same characteristic times as Tables 4 and 5,
but in units of the age of the universe in each cosmological model.
With the exception of the $\Omega =0, \, \Lambda =1$ model, 
all ratios in
Table 6 are close to 0.30$\pm 0.02$, implying that the characteristic 
evolution
time is about equal to 1/(3.2$\pm 0.2$) of the age of the universe, 
whatever the cosmological model, in PWLE models. In
LEXP models, this ratio is 1/(6.7$\pm $0.9). 
The characteristic times in Tables 4 and 5, apparently quite
dependent of the cosmology, are indeed surprinsingly
constant when expressed in terms of the age of the universe. 

\section{An algorithm for producing quasar catalogues}
\medskip
\subsection{ The algorithm}
Fixing the evolutionary law, the cosmology (including 
the choice of $H_0$) and
the catalogue limits (i.e. $m_1,m_2,z_1,z_2$ 
which allow the box $\cal B$ and domains $\cal D$ and $\cal D^*$
to be defined), each QSO $(m,z)$ is fully
determined by the other pair of variables $(V/V_{a},M_0)$. 
This is easy to understand from Fig. 1, where the Mattig functions appear.
All of our previous discussion shows that under the PLE hypothesis
(which was checked on the data), any sample is fully determined from 
its limiting magnitudes and redshifts, the GLF
and a uniform distribution of the $V/V_{a}$. From the probability 
distribution function (PDF) of absolute magnitudes $M_0$,
which is $\phi (M_0) \times V_{a}(M_0)$, 
and from the $V/V_{a}(M_0)$ ratios uniformly spread over $[0,1]$, it
is possible to draw Monte--Carlo catalogues.

The algorithm is shown in Fig. 4. 
Choosing the cosmology and a functional form for the evolution, the first
part of the work is to compute the evolutionary parameter
from the constraint $<V/V_{a}>=0.5$, using the Boyle sample.
At the same time, $V_{a}(M_0)$ is computed and used to obtain the GLF.
In the second part,
two series of random numbers, $\epsilon_1$ and $\epsilon_2$, are used to
extract the $V/V_{a}$ and the absolute magnitude $M_0$ of each object in
the synthetic catalogue from the relations:

\EQA
\epsilon_1&=&V/V_{a}={V-V_{sup}-\sum V_{out}\over V_{inf}-V_{sup}-\sum
V_{out}},\nonumber\cr
\epsilon_2&=&{\int_{-\infty}^{M_0} dM \phi (M) V_{a} (M) \over 
\int_{-\infty}^{+\infty}dM \phi (M) V_{a} (M) }.
\label{catapar}
\ENA

Obtaining the redshifts and apparent magnitudes is straightforward from a
simple inversion procedure. Synthetic catalogues have been produced in
the past, where both a LF {\it plus} a redshift
histogram were jointly fitted. As already explained, this
method suffers from at least two problems: 
Essentially, the two distributions may
be incompatible in different cosmologies.
Our method has the advantage of avoiding these problems since we use only one
distribution, the GLF, and the other distributions (redshift, magnitude)
are derived only after the cosmology and the evolution have been fixed.
The algorithm shown in Fig. 4 ensures the coherence of these
distributions, moreover it takes the evolution into account very 
accurately.
This is not obviously true in previous methods where it is not clear how 
the evolution intervenes. In fact, we used the Cosmological Principle 
to replace the redshift
distribution of the standard method by demanding the uniformity of $V/V_{a}$.
Moreover, the limiting magnitude and the redshift range for the synthetic 
sample may be different from their values in the input sample.

\begin{figure}
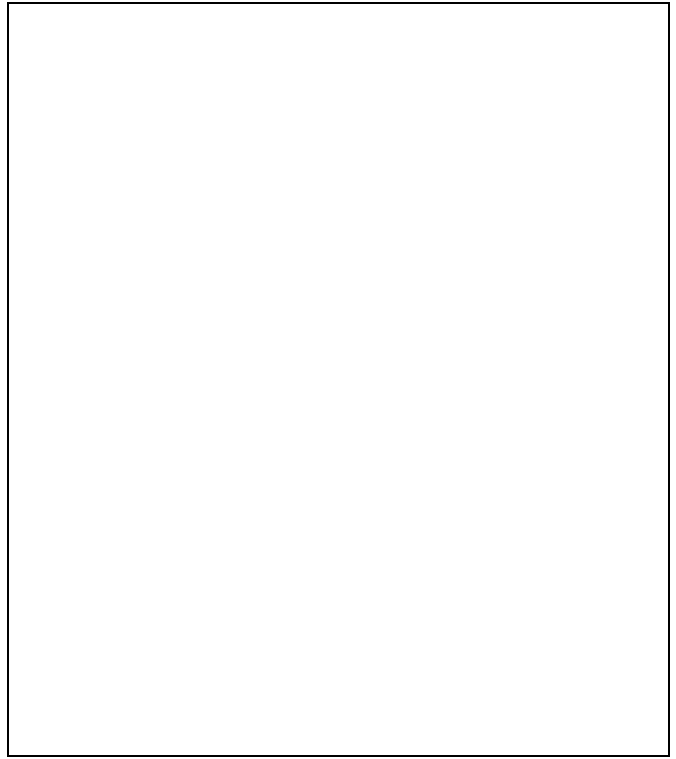

\picplace{10cm}
\caption{Algorithm for the construction of synthetic catalogues}
\end{figure}

\subsection{ Redshift and Magnitude Histogrames}
The input \cite{boyle90} and output (synthetic catalogue with the
same observational biases) redshift histogrames are compared in Fig.
5 for models $b$ and $f$ of Table 1.
Similarly, the histogrames of absolute magnitude are compared in Fig. 6.
The fit is quite satisfactory for both models; this is true for all models
we tried.
It is possible to introduce various biases in the synthetic catalogues,
depending on which sort of catalogue we want to simulate. For example, for
UVX quasars it is well known that the color selection criteria introduce
a deficiency of QSOs in the redshift range $[1.5,1.8]$ \cite{boyle88}.
Moreover, QSOs are
variable objects, which can be simulated by randomly selecting a magnitude 
variation
in a pre--defined distribution function. Noise in the magnitude and redshift 
measurements may be simulated in the same way.

\begin{figure}
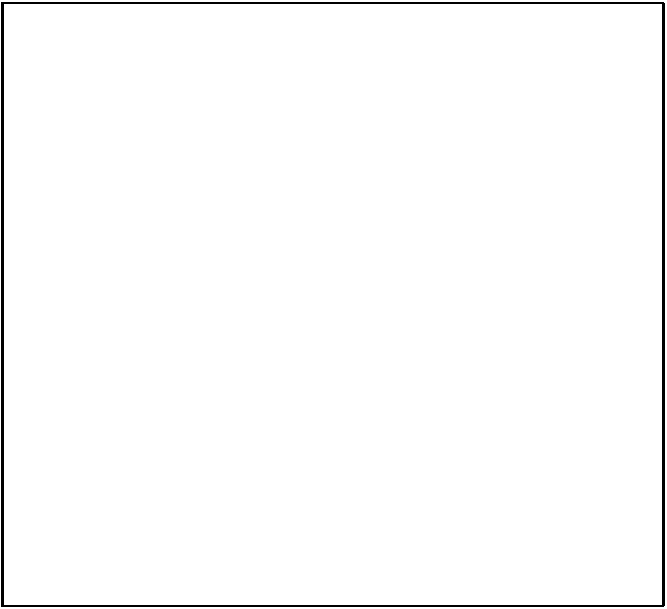

\picplace{8cm}
\caption{Redshift Histogrames: (1): Boyle sample; (2): simulated
catalogue, translated 15 upwards, model $b$; (3): idem, translated 35 upwards, 
model $f$ }
\end{figure}

\begin{figure}
\picplace{8cm}
\caption{Absolute Magnitude Histogrames: dashed curve: Boyle sample; 
full line: simulated catalogue, model $b$}
\end{figure}

\begin{figure}
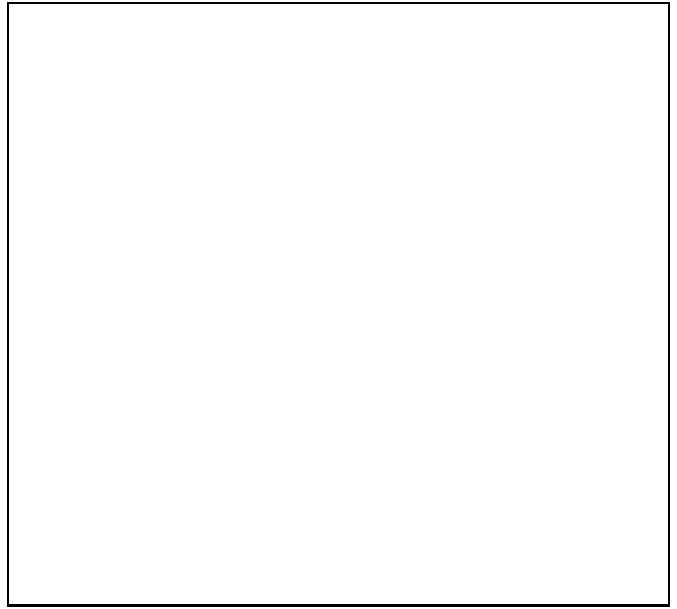

\picplace{8cm}
\caption{Absolute Magnitude Histogrames: dashed curve: Boyle sample; 
full line: simulated catalogue, model $f$}
\end{figure}

\section{Conclusion}

We have reviewed in detail the $V/V_{max}$ computation in various cosmologies.
The Mattig relation is used to understand how to compute the available volume
$V_a$ of a QSO, even in complicated cases. Both the Pure Luminosity
Evolution (PLE) assumption and the Cosmological Principle allow
the variables $(M_0,V/V_{a})$ to be used instead of $(m,z)$ to construct
synthetic QSO samples in a coherent way.
Some of the previous catalogue constructions were shown to suffer problems.
The Global Luminosity Function used for this purpose is derived in a 
slightly different way from previous calculations
since we fixed the evolution parameter by the condition
$<V/V_{a}>=1/2$. This significantly changes the value of this parameter,
and hence of the GLF. 

A surprising result is that the QSO characteristic evolution times
are constant when expressed in terms of the age of 
the universe, regardless of the cosmology.
The PLE hypothesis was checked on the data, which comfirms previous results 
\cite{boyle88}. The advantage of such catalogues is that we tightly control all 
aspects of the sample: 
the parent cosmology, the strength and the functional form of the
evolution, the luminosity function, the magnitude and redshift depths, 
and the effects of magnitude and redshift biases, all of which influence
the results of the new cosmological test introduced in Paper I. In Paper
I we apply the 
test to catalogues with different redshift limitations, which
leads us to propose a new observational strategy for the construction of
future QSO samples.
It is possible to test a catalogue with a wrong hypothesis, for 
example on the evolution, by assuming a power--law evolution and running the 
cosmological test with the exponential hypothesis.
Moreover, our method is tractable with a Pure Density Evolution hypothesis,
provided that density--weighted volume elements $\rho dV$ are substituted
for all volume elements $dV$.

\begin{table*}
\caption[]{%%
The LF (Eq. (\ref{fctlum})) in various redshift ranges
for four arbitrary cosmologies and two evolution laws.
Since we are interested in input parameters to construct catalogues,
confidence intervals are not needed. Only the best
fit parameters are given. We choose $H_0=50 km s^{-1} Mpc^{-1}$.%% 
}
\begin{flushleft}
\begin{tabular}{ccccccccccccccc}
\noalign{\smallskip}
\hline
\noalign{\smallskip}
&&&&&&&&&&&\cr
&$\Omega $&$\Lambda $&\hfil  $k_L$&\hfil $\alpha$\hfil &\hfil $\beta$\hfil%%
 &\hfil  $M_{\star}$& $\phi^{\star}$&\hfil $P(>\chi^2)$\hfil%%
 &\hfil $P(>\chi^2)$\hfil &\hfil $P(>\chi^2)$\hfil &\hfil  $P(>\chi^2)$%%
 \hfil  \cr
%%&$\Omega $&$\Lambda $&\hfil  $k_L$&\hfil $\alpha$\hfil &\hfil $\beta$\hfil%%
%% &\hfil  $M_{\star}$& $\phi^{\star}(Mpc^{-3})$& \multicolumn{4}{1} \hfil $P(>\%%chi^2)$\hfil \cr
%%\cline{9-12}
& &&&&&&$10^{-6} Mpc^{-3}$&$z_1=$0.3 & 0.3&1.0&1.7 \cr
& &&&&&&&$z_2=$2.2 & 1.0&1.7 &2.2\cr
\hline\noalign{\smallskip}
PWLE&&&&&&&&&&\cr
a&0&0&2.94&-3.32&-1.38&-23.41&1.8$10^{-6}$&\hfil 0.97\hfil &\hfil 0.20\hfil &\hfil 0.78\hfil &\hfil 0.80\hfil \cr
b&1&0&3.60&-3.89&-1.30&-22.10&1.1$10^{-5}$&\hfil 0.99\hfil &\hfil 0.99\hfil &\hfil 0.69\hfil &\hfil 0.92\hfil \cr
c&0&1&1.91&-2.43&-1.23&-23.76&2.6$10^{-6}$&\hfil 0.89\hfil &\hfil 0.15\hfil &\hfil 0.88\hfil &\hfil 0.58\hfil \cr
d&1&1&3.81&-4.22&-1.37&-22.25&3.2$10^{-6}$&\hfil 0.99\hfil &\hfil 0.92\hfil &\hfil 0.66\hfil &\hfil 0.81\hfil \cr
\noalign{\smallskip}
\hline
\noalign{\smallskip}
LEXP&&&&&&&&&&\cr
e&0&0&6.65&-3.51&-1.49&-22.18&1.5$10^{-6}$&\hfil 0.99\hfil &\hfil 0.58\hfil &\hfil 0.78\hfil &\hfil 0.40\hfil \cr
f&1&0&11.8&-4.61&-1.55&-19.84&6.3$10^{-6}$&\hfil 0.97\hfil &\hfil 0.88\hfil &\hfil 0.57\hfil &\hfil 0.24\hfil \cr
g&0&1&1.90&-2.69&-1.49&-24.66&1.2$10^{-6}$&\hfil 0.76\hfil &\hfil 0.16\hfil &\hfil 0.23\hfil &\hfil 0.58\hfil \cr
h&1&1&9.71&-4.62&-1.60&-20.13&2.1$10^{-6}$&\hfil 0.99\hfil &\hfil 0.64\hfil &\hfil 0.57\hfil &\hfil 0.01\hfil \cr
%&&&&&&&&&&&&\cr
\noalign{\smallskip}
\hline
\end{tabular}
\end{flushleft}
\end{table*}

\begin{table}
\caption[]{%%
Power Law Luminosity Evolution (PWLE) parameter in different cosmologies.
The model $\Omega =0, \, \Lambda =1$, which is marked with an
asterisk, is close to a universe without a Big--Bang, and not 
very reliable.%
}
\begin{flushleft}
\begin{tabular}{lllllllllllllll}
\noalign{\smallskip}
\hline
\noalign{\smallskip}
$\Lambda $ &$\Omega $&  0.0   &   0.2  &    0.4   &   0.6  &    0.8  &    1.0  \cr
0.0 && 2.94 &    3.16  &   3.28  &   3.44  &   3.50  &   3.60 \cr
0.2 && 2.88 &    3.13  &   3.28  &   3.47  &   3.50  &   3.63 \cr
0.4 && 2.78 &    3.10  &   3.28  &   3.47  &   3.53  &   3.66 \cr
0.6 && 2.66 &    3.06  &   3.28  &   3.47  &   3.60  &   3.69 \cr
0.8 && 2.47 &    3.00  &   3.31  &   3.47  &   3.63  &   3.75 \cr
1.0 && 1.91(*) &    2.94  &   3.35  &   3.50  &   3.66  &   3.81 \cr
%&&&&&&\cr
\noalign{\smallskip}
\hline
\end{tabular}
\end{flushleft}
\end{table}

\begin{table}
\caption[]{%%
Same as Table 2 for LEXP, Exponential Luminosity Evolution.}
\begin{flushleft}
\begin{tabular}{lllllllllllllll}
\noalign{\smallskip}
\hline
\noalign{\smallskip}
$\Lambda $ &$\Omega $& 0.0 & 0.2 & 0.4 & 0.6 & 0.8 &    1.0 \cr       
0.0 && 6.65  &   8.02  &   9.15  &  10.10 &   11.02 &   11.77 \cr
0.2 && 6.02  &   7.40  &   8.65  &   9.71 &   10.59 &   11.40 \cr
0.4 && 5.28  &   6.84  &   8.15  &   9.28 &   10.15 &   11.03 \cr
0.6 && 4.40  &   6.15  &   7.53  &   8.78 &    9.71 &   10.65 \cr
0.8 && 3.40  &   5.40  &   6.96  &   8.21 &    9.28 &   10.15 \cr
1.0 && 1.90(*)  &   4.59  &   6.28  &   7.65 &    8.78 &    9.71 \cr
%&&&&&\cr
\noalign{\smallskip}
\hline
\end{tabular}
\end{flushleft}
\end{table}

%\medskip
\begin{table}
\caption[]{%%
Evolution characteristic times: look--back time to which
all luminosities were higher by a factor $e$, given 
the PWLE parameter in Table 2. Times are in
units of $H_0^{-1}$, the mean is 0.27 and the dispersion is 0.04. 
}
\begin{flushleft}
\begin{tabular}{lllllllllllllll}
\noalign{\smallskip}
\hline
\noalign{\smallskip}
$\Lambda $ &$\Omega $&  0.0   &   0.2  &    0.4   &   0.6  &    0.8 &
  1.0 \cr       
0.0 && .29  &    .27   &   .26   &   .24   &   .23   &   .23  \cr
0.2 && .30  &    .28   &   .26   &   .24   &   .24   &   .23  \cr
0.4 && .32  &    .29   &   .27   &   .25   &   .24   &   .23  \cr
0.6 && .35  &    .30   &   .28   &   .25   &   .24   &   .24  \cr
0.8 && .39  &    .31   &   .28   &   .26   &   .25   &   .23  \cr
1.0 && .52(*)  &    .33   &   .29   &   .27   &   .25   &   .24  \cr
%&&&&&&\cr
\noalign{\smallskip}
\hline
\end{tabular}
\end{flushleft}
\end{table}

%\medskip
%\medskip
\begin{table}
\caption[]{%%
Same as Table 4, but for LEXP. Mean characteristic time:
0.13; dispersion: 0.05. %%
}
\begin{flushleft}
\begin{tabular}{lllllllllllllll}
\noalign{\smallskip}
\hline
\noalign{\smallskip}
$\Lambda $ &$\Omega $&  0.0   &   0.2  &    0.4   &   0.6  &    0.8
&    1.0
  \cr
0.0 && .15  &    .12  &    .11  &    .10  &    .09  &    .08  \cr
0.2 && .17  &    .13  &    .12  &    .10  &    .09  &    .09  \cr
0.4 && .19  &    .15  &    .12  &    .11  &    .10  &    .09  \cr
0.6 && .23  &    .16  &    .13  &    .11  &    .10  &    .09  \cr
0.8 && .29  &    .18  &    .14  &    .12  &    .11  &    .10  \cr
1.0 && .53(*)  &    .22  &    .16  &    .13  &    .11  &    .10  \cr
%&&&&&&\cr
\noalign{\smallskip}
\hline
\end{tabular}
\end{flushleft}
\end{table}
%\medskip
\begin{table}
\caption[]{%%
PWLE characteristic times (Table 4) in units of the age of 
the universe in  the
corresponding cosmology. Mean: 0.31; dispersion: 0.02. These
ratios are far more independent of cosmology than the times in Table 4.
}
\begin{flushleft}
\begin{tabular}{lllllllllllllll}
\noalign{\smallskip}
\hline
\noalign{\smallskip}
$\Lambda $ &$\Omega $&  0.0   &   0.2  &    0.4   &   0.6  &    0.8 &
  1.0
  \cr
0.0   && .29   &   .32  &    .33  &    .33   &   .34  &    .34  \cr
0.2   && .28   &   .31  &    .33  &    .32   &   .33  &    .34  \cr
0.4   && .28   &   .31  &    .32  &    .32   &   .33  &    .33  \cr
0.6   && .27   &   .30  &    .31  &    .31   &   .32  &    .32  \cr
0.8   && .25   &   .29  &    .30  &    .30   &   .31  &    .31  \cr
1.0   && .13(*)&   .28  &    .29  &    .30   &   .30  &    .31  \cr
%&&&&&&\cr
\noalign{\smallskip}
\hline
\end{tabular}
\end{flushleft}
\end{table}
%\medskip
\begin{table}
\caption[]{%%
Same as Table 6, but for LEXP. Mean: 0.15; dispersion:
0.02. As for PWLE, the characteristic time depends 
far less on the age of the universe than on the Hubble time.
}
\begin{flushleft}
\begin{tabular}{lllllllllllllll}
\noalign{\smallskip}
\hline
\noalign{\smallskip}
$\Lambda $ &$\Omega $&  0.0   &   0.2  &    0.4   &   0.6  &    0.8 &
  1.0
  \cr
0.0 &&  .15   &   .15  &    .14   &   .14  &    .13  &    .13  \cr
0.2 &&  .16   &   .15  &    .14   &   .14  &    .13  &    .13  \cr
0.4 &&  .16   &   .16  &    .15   &   .14  &    .13  &    .13  \cr
0.6 &&  .17   &   .16  &    .15   &   .14  &    .13  &    .13  \cr
0.8 &&  .19   &   .17  &    .15   &   .14  &    .13  &    .13  \cr
1.0 &&  .13(*)&   .18  &    .16   &   .14  &    .14  &    .13  \cr
%&&&&&&\cr
\noalign{\smallskip}
\hline
\end{tabular}
\end{flushleft}
\end{table}
%\medskip

\paragraph{Acknowledgements}

We thank B. Fort, P.Y. Longarreti, G. Soucail, J.P. Picat, R. Pell\'o 
and J.F. Leborgne for support and discussions, and especially Jim
Bartlett for a careful reading of the manuscript and enlighting
suggestions.
L.V.W. thanks the french MESR for grant 93135.
This work was supported by grants from the french CNRS (GdR Cosmologie)
 and from the
European Community (Human Capital and Mobility ERBCHRXCT920001).

\bibliography{cosmo3}{}
\bibliographystyle{astron}

\end{document}